\documentclass[aps,prl,twocolumn,showpacs,preprintnumbers]{revtex4}

\usepackage[dvips]{graphicx}
\usepackage{latexsym}
\usepackage{graphicx}
\usepackage{epsfig}
\usepackage{graphicx,graphics,epsfig,epstopdf,subfigure,color,psfrag}
\usepackage[english]{babel}
\usepackage[latin1] {inputenc}
\usepackage{amsmath,amsfonts,amssymb}
\usepackage{bm}
\usepackage{color}

\newcommand \be {\begin{equation}}
\newcommand \bea {\begin{eqnarray}}
\newcommand \ee {\end{equation}}
\newcommand \eea {\end{eqnarray}}
\newcommand \bed {\begin{displaymath}}
\newcommand \eed {\end{displaymath}}

\newcommand{\bit}{\begin{itemize}}
\newcommand{\eit}{\end{itemize}}

\newcommand{\bgar}{\begin{eqnarray}}
\newcommand{\enar}{\end{eqnarray}}

\begin{document}
\title{Nuclear Shadowing in   the Holographic Framework}
\author{L. Agozzino$^{a,b}$, P. Castorina$^{a,b}$, and P. Colangelo$^{c}$}
\affiliation{$^{a}$Dipartimento di Fisica, Universit\'a  di Catania, via S. Sofia 62, 95125 Catania, Italy\\
$^{b}$INFN, Sezione di Catania, via S. Sofia 62, 95125 Catania, Italy\\
$^{c}$INFN, Sezione di Bari, via Orabona 4, 70126 Bari, Italy}
 
\begin{abstract}
The nucleon structure function $F_2^N$ computed in a holographic framework can be used to describe nuclear deep inelastic scattering effects provided that a rescaling of the $Q^2$ momentum and of the IR hard-wall parameter $z_0$ is made. The ratios $R_A=F_2^A/F_2^N$ can be obtained in terms of a single rescaling parameter $\lambda_A$ for each nucleus. The resulting ratios agree with experiment in a wide range of the shadowing region.

\end{abstract}
\maketitle

\section{Introduction}

The AdS/CFT correspondence \cite{adscft}, an important tool to analyze nonperturbative aspects of gauge theories, has been successfully used to study  features of QCD  \cite{finite}. In its application to  deep inelastic scattering (DIS) at strong coupling \cite{Polchinski,Brower:2010wf},
the nucleon structure function  $F_2^N(x,Q^2)$ at small   Bjorken variable  $x$
has been represented as the sum of a conformal term and of a contribution due to quark confinement,  crucial to  fit the data. The evaluation of these contributions requires the holographic nucleon wave function, which is assumed to be peaked at distances $O(1/Q^\prime)$ close to the infrared boundary $z_0$, with $Q^\prime$ of the order of the nucleon mass.

Nonperturbative, confinement dynamics shows up also  in the modification of the structure functions of a nucleon in a nucleus. The proposed theoretical models of these effects are  based on the effective change of the mean square distances among quarks and gluons in a nuclear environment with respect to the free nucleon case \cite{arneodo}.

Nuclear effects are described  comparing the structure functions of the nuclear target per nucleon  to the free nucleon ones and, for electroproduction, the ratios 
$R_A= F_2^A(x,Q^2)/F_2^D(x,Q^2)$ are considered, with $F_2^A$ and $F_2^D$  the structure functions per nucleon in the nucleus $A$ and in deuterium $D$ (where the nuclear binding is considered negligible). Nuclear modifications depend on  $x$:  $x \le 0.1$  is the shadowing region where $R_A < 1$; for large $x$ one has the so-called EMC effect; in the range $0.1 < x< 0.25$ there is the anti-shadowing with $R_A > 1$, usually obtained  by the energy-momentum sum rule.
The understanding of these modifications requires to evaluate the  \textquotedblleft distortion\textquotedblright  of the free nucleon wave function due to nuclear binding. In the small-$x$ region, dominated by Pomeron exchange,  the AdS/CFT strong coupling BPST Pomeron kernel  \cite{Brower:2006ea} is a good framework to study nuclear deep inelastic structure functions,  provided one knows the holographic baryon wave function in the nucleus. 
A simplified approach can be based on the observation that the spatial separation between quarks determines the strength of the quark-Pomeron coupling \cite{pov} and that the effective confinement size is modified in a nucleus. 

Following this point of view,  we  consider an approach to shadowing in an AdS/CFT framework, describing the nuclear binding effects on the nucleon wave function through an effective distance $1/Q^\prime_A$ and an effective confinement boundary $z_0^A$,   and  studying the scaling properties of the  holographic structure function $F_2$ under the change $Q^\prime \rightarrow  Q^\prime_A $ and $z_0 \rightarrow z_0^A$. In this way, nuclear effects are described  by rescaling the confinement parameters. The  
results  agree with  the experimental data.

\section{Approaches to nuclear DIS effects}

In models of shadowing and EMC effects,  the  description of the nuclear modifications is  based on the change of the effective
mean square distance among quarks and gluons in a nuclear environment with respect to the free nucleon case \cite{arneodo}.
  
In the $x$-rescaling model the EMC effect is described by rescaling  $x$  in the free nucleon structure function  \cite{aku},
\be
F_2^A(x,Q^2)=F_2^D(x/\hat z,Q^2) \,\,\, ,
\ee
with $\hat z \simeq 1- \epsilon/M$, $M$  the proton mass and $\epsilon$  the energy necessary to \textquotedblleft ionize\textquotedblright  a nucleus and make it emit a nucleon. However, the values of  $\epsilon$ able to  fit the large-$x$ data  exceed the nuclear binding calculations \cite{arneodo}.

The  $Q^2$-rescaling model of EMC effect is based  on the relation \cite{close,close2}
\be
F_2^A(x,Q^2)=F_2^D(x, \chi_A Q^2) \,\,\, , 
\ee
indicating that the effective $Q^2$ for a bound nucleon is different from the free nucleon. This dynamical property  is investigated here in the holographic framework. It is related to the modification of the quark confinement size in the nucleus \cite{close,close2}:
quarks and gluons are no longer confined to specific nucleons, but  spread over distances larger  than the  free nucleon size. By studying the structure function moments,  starting from a $Q^2$ region where the valence picture is a good approximation,  one can show that in QCD, for large $Q^2$, this change of scale is connected to  the strong coupling constant $\alpha_s$.  The $x$- and $Q^2$-rescaling models, although different in their starting points, can be  related \cite{close4}.

The QCD $Q^2$ dependence of structure functions can be also applied to shadowing at small $x$,  including the effect of gluon recombination in nuclei which is neglected in the free nucleon evolution equation \cite{qiu}. 
Modifying the linear $Q^2$ evolution equation one  shows  that the recombination depletes the gluon distribution at small $x$,  which reflects into a depletion of sea quark distribution \cite{qiu}. The $Q^2$ dependence of parton distributions in nuclei based on linear QCD evolution equations at next-to-leading order is described in  ref.\cite{eskola1}.

A different nonperturbative approach considers that the small-$x$ behavior of  $F_2$ is controlled by Pomeron exchange \cite{castorina}. In a nuclear environment the effective coupling of the Pomeron to a quark is suppressed because of the nucleon overlap. Although quarks and gluons are no longer confined to specific nucleons but rather spread on distances larger than the  free nucleon size, the average spatial separation between quarks before color neutralization decreases,
with  the Pomeron coupling  directly related with this typical size \cite{pov}.

Hence, a physical description of EMC and shadowing effects can be based on the effective modification of the dynamical scales in deep inelastic scattering on nuclear target  respect to the free nucleon case.
 This rescaling, and in particular the $Q^2$ one,  is a property of the AdS/CFT approach to deep inelastic scattering, not only in the conformal limit but also if  the confinement dynamics is taken into account. 

\section{Nuclear Structure Functions in Holographic Framework}

The possibility of an   approach to  DIS on a proton based on AdS/CFT duality was analized in the Polchinski-Strassler first proposal  \cite{Polchinski}. Here we adopt the method in  \cite{Brower:2010wf},  based on the calculation of the virtual $\gamma^* p$ total cross section, which allows to  express, e.g., the structure function $F_2$ as the sum of two contributions: a model-independent  term for conformal gauge theories and an additional non-conformal term accounting for confinement. This latter (model dependent) contribution is obtained by breaking conformal invariance through a sharp cut-off (\textquotedblleft hard-wall\textquotedblright ) of the AdS  holographic space  \cite{others}.

One starts from the matrix element of two 
 R-currents  
in a hadron of momentum $P$ and charge $\cal Q$,

\bea
T^{\mu \nu} &\equiv& i \int d^4y e^{i q \cdot y}  \langle P {\cal Q} | T[J^\mu(y) J^\nu(0)] | P {\cal Q} \rangle \nonumber \\
 &=&F_1(x,Q^2)(\eta^{\mu \nu}-\frac{q^\mu q^\nu}{q^2}) \nonumber \\
&+& \frac{2 x}{q^2}F_2(x,Q^2)(P^\mu+\frac{q^\mu}{2x})(P^\nu+\frac{q^\nu}{2x}) 
\,\,\, \label{eq:Tmunu}
\eea
(with $\mu,\nu$ four-dimensional indices, $\eta^{\mu \nu}$ Minkowski metric,  $x=Q^2/2 P \cdot q$ and $Q^2=-q^2$)
 which allows to extract the DIS structure functions for electron-hadron  scattering and, in particular, $F_2(x,Q^2)$.
The AdS/CFT calculation involves 
 the couplings 
\be
g_s=\frac{g_{YM}^2}{4 \pi}=\alpha_{YM}=\frac{\lambda}{4 \pi N_C},  \,\,\,\,  R=\alpha^{\prime \frac{1}{2}} \lambda^{\frac{1}{4}}
\ee
with $g_s<<1$ and $\lambda >>1$. $R$ is the  AdS radius. 
In the following,  the coupling  $\displaystyle \rho=2/\sqrt{\lambda}$ is  used.

The dual string calculation of the matrix element (\ref{eq:Tmunu}), or of its imaginary part  appearing in DIS processes, describes the scattering   in the AdS space, and
 involves various quantities. First,  to describe the transition $\gamma^*N \to \gamma^* N \equiv 1,2 \to 3,4$,
states dual to the initial-final nucleon $N$ are required,  i.e. the hadronic  state $|PQ \rangle$ in (\ref{eq:Tmunu}). These states are represented by normalizable wave functions $\phi^N(z)$ depending on the holographic coordinate $z$ (positive and with the UV brane corresponding to $z=0$),  and are  obtained in principle from a
 suitable equation of motion.
The calculation of the matrix element (\ref{eq:Tmunu}) involves the transition  function 
\be 
P_{24}(z)=\sqrt{-g} \, \left(\frac{z}{R}\right)^2 \, \phi^N(z) \phi^N(z) \,\,\, . \label{P24}
\ee 

The current that couples to the hadrons in  (\ref{eq:Tmunu}) induces non-normalizable modes of the gauge fields. In the bulk, such fields $\cal A$ satisfy Maxwell's equations of motion; their solutions, in the Lorentz gauge and for $R=1$,  are given in terms of Bessel functions:  ${\cal A}_\mu(y,z) =  n_\mu (Qz) K_1(Qz) e^{i q\cdot y}$ and ${\cal A}_z(y,z) = i (q \cdot n) (Qz) K_0(Qz) e^{i q\cdot y}$, with $n_\mu$ a  polarization vector.
To  determine    the structure function $F_2$ in   (\ref{eq:Tmunu})
a transition function $P_{13}$ is  needed,  and is  given by \cite{Brower:2010wf,Pire:2008zf}
\be
P_{13}(z,Q^2) = \frac{1}{z} (Qz)^2 [K_0^2(Qz) + K_1^2(Qz)] \label{P13}
\ee
with $Q=\sqrt{Q^2}$.

The last ingredient is the scattering kernel. This has been expressed in terms of a Pomeron Regge pole contribution  \cite{Brower:2006ea}, and allows to write the structure function $F_2$ at low-$x$ as an eikonal sum with a convolution of the transition functions \eqref{P24}  and \eqref{P13}  \cite{Brower:2010wf}:
\bea
F_2^N(x,Q^2) &=& \frac{Q^2}{2 \pi^2}
 \int d^2 b \int dz dz^\prime P_{13}(z,Q^2) P_{24}(z') \nonumber \\ &\times &{\rm Re} \left( 1- e^{i \chi (s,b,z,z^\prime)}\right) \,\,\, . \label{F2N}
\eea
$b$  is the impact parameter,  with $\vec b$  the transverse Minkowski space vector for  $\gamma^* p$ scattering; 
 $s$  is the center-of-mass energy squared of the $\gamma^*$-target system.
The eikonal $\chi$ can be derived for conformal theories;  it can also be modified by the inclusion of conformal symmetry-violating  effects.

\subsection{Conformal limit}

For conformal fields the free nucleon structure function $F_2^N $ can be obtained from (\ref{F2N}) and is given by \cite{Brower:2010wf}
\bea
F_2^N(x,Q^2) &=& \frac{g_0^2 \rho^{3/2}}{32 \pi^{5/2}}   \int dz dz^\prime \frac{z z^\prime Q^2}{\tau^{1/2}} P_{13}(z,Q^2 ) P_{24}(z^\prime) \nonumber \\
&\times& e^{(1-\rho) \tau} \exp{[\Phi(z,z{'},\tau)]} \,\,\, , \label{F2Nconf}
\eea
where $x \simeq Q^2/s$ and $g_0^2$ a constant. The conformal invariant $\tau$ is defined as  $\tau = \log{(\rho z z^\prime s/2)}$.
 The function $\Phi$ is the BPTS  Pomeron kernel integrated in impact parameter   \cite{Brower:2006ea}, 
\be
\Phi(z,z^\prime,\tau) = - \frac{(\log{z} - \log{z^\prime})^2}{ \rho \tau }\,\,\, .
\ee
The transition function $P_{24}$ involves the nucleon  wave function in the bulk $\phi^N(z)$.
 In ref.\cite{Brower:2010wf} it is assumed that the wave function $\phi^N(z)$ is sharply peaked near the infrared boundary $z_0$, with $1/Q^\prime \leq z_0$ and $Q^\prime$ close to the nucleon mass:
\be
P_{24}(z^\prime) \simeq \delta(z^\prime - 1/Q^\prime) \,\,\, , \label{localP24}
\ee
an expression adopted in the following. An explicit bulk model for the nucleon would be required to improve the wave function profile; modifications of this local approximation can be considered \cite{next}.
A further simplification  consists in replacing also $P_{13}$ by a local expression,
\be
P_{13}(z, Q^2) \simeq C \delta(z - 1/Q) \label{localP13}
\ee
with $C \simeq 1$. Since the  integrand with  $P_{13}$ in Eq.(\ref{F2Nconf}) is peaked for $z \simeq 1/Q$, one can  verify that this is a good approximation of Eq.(\ref{P13}).
The resulting  $F_2$ reads \cite{Brower:2010wf}:
\be
F_2^N(x,Q^2) = \frac{g_0^2 \rho^{3/2}}{32 \pi^{5/2}} \frac{Q}{Q^\prime} \frac{1}{\tau^{1/2}} e^{(1-\rho) \tau} 
e^{-[log^2(Q/Q^\prime)/\rho \tau]} \,\,\, . \label{F2Nconf1}
\ee

The nucleon structure function $F_2^A$ at small $x$ in a nuclear environment can be  obtained rescaling the effective size of the nucleon wave function in the nucleus $A$
\be\label{Qresc}
Q^\prime_A = \lambda_A  Q^\prime
\ee
in Eq.(\ref{F2Nconf1}). In the conformal limit $F_2$ depends on the ratio $Q/Q^\prime$ and therefore the rescaling 
$Q^\prime_A \rightarrow  Q^\prime$ corresponds to the $Q^2$ rescaling $Q^2 \rightarrow Q^2/\lambda_A^2$.  Hence,  in this limit one has
 $R_A = F_2^A/F_2^D$ neglecting  the proton-neutron difference.
Therefore,  in the conformal limit the $Q^2$-rescaling at small $x$ naturally arises in the AdS/CFT  approach. This  is not surprising, since the limit is reliable at large $Q^2$.
Notice that in the local approximation (\ref{localP24},\ref{localP13}), $\tau = \log{(\rho Q / 2 x Q^\prime)}$, therefore
the rescaling $Q^\prime_A = \lambda_A  Q^\prime$ could be reabsorbed in  $x \rightarrow \lambda_A x$. However, due to the $Q^2$ dependence of $F_2$ in Eq.(\ref{F2Nconf1}), the $x$-rescaling is not completely equivalent to the rescaling in $Q^2$,  and  $F_2^A(x,Q^2/\lambda^2) \neq F_2^A(\lambda x, Q^2)$. The $x$-rescaling method in the  holographic framework will be discussed in a dedicated study \cite{next}.

\subsection{Confinement effects}

The expression for $F_2^N$ for the free proton structure functions, based on the conformal BPST Pomeron,  does not  fit the HERA data in the low $Q^2$ region, where confinement is the main dynamical mechanisms \cite{Brower:2010wf}. One needs to account for  confinement,  which can be described  including an infrared boundary  $z_0$ on the $z$ coordinate of the bulk.
This scale could be related to the $\Lambda_{QCD}$  parameter.
It produces a mass gap, a modification of the eikonal and a non-conformal contribution to $F_2^N$ which, for a single Pomeron,  reads \cite{Brower:2010wf}:
\bea
F_{2ct}^N(x,Q^2,z_0) &=& \frac{g_0^2 \rho^{3/2}}{32 \pi^{5/2}} \int dz dz^\prime \frac{z z^{\prime}Q^2}{\tau^{1/2}} P_{13}(z,Q^2 ) P_{24}(z^\prime) \nonumber \\
&\times& e^{(1-\rho) \tau} \,\, e^{- \frac{\log^2{(z z^\prime/z_0^2)}}{ \rho \tau}}\,\, G(z,z^\prime,\tau) . \label{F2Nnew}
\eea
In this expression the  $z_0$ dependence is shown explicitly; $G(z,z^\prime,\tau)$ is given by
\be
G(z,z^\prime,\tau) = 1- 2 \sqrt{\rho \pi \tau} e^{\eta^2} erfc(\eta),
\ee
and
\be
\eta = \frac{-\log{(zz^\prime/z_0^2)} + \rho \tau}{ \sqrt{\rho \tau} } \,\,\, .
\ee
With the local approximation (\ref{localP24},\ref{localP13}),  Eq.(\ref{F2Nnew}) reduces to
\bea
F_2^N(x,Q^2,Q_0^2) &=& \frac{g_0^2 \rho^{3/2}}{32 \pi^{5/2}} \frac{(Q/Q^\prime)}{\tau^{1/2}} e^{(1-\rho) \tau} \nonumber \\  
&\times& e^{- \frac{\log^2{(Q_0^2/(Q Q^\prime))}}{ \rho \tau}} G(\frac{1}{Q},\frac{1}{Q^\prime},\tau) \,\,\, ,
\label{F2NQ0}
\eea
where $Q_0=1/z_0$ \cite{Brower:2010wf}.

In the description  of nuclear effects by the rescaling $Q^\prime_A = \lambda_A  Q^\prime$,  the  $\tau$ dependence on the ratio $Q/Q^\prime$  is remarkable, and  would suggest a $Q^2$-rescaling.  However, there is also a nontrivial $Q^2$ behavior in the log-factors and in 
$\eta$  due to the new scale $Q_0$. 
The dependence on $Q_0$ in Eq.(\ref{F2Nnew}) is in the form $Q_0^2/QQ^\prime$;  therefore,  the rescaling $Q^\prime_A = \lambda_A  Q^\prime$ can be reabsorbed in the  $Q^2$ rescaling  $Q^2 \rightarrow Q^2/\lambda_A^2$ provided that
the confinement distance in nuclear environment scales in the same way,  
\be\label{Q0resc}
Q_0^2 \rightarrow Q_0^2/\lambda_A^2 \,\,\,.
\ee
Our phenomenological analysis  is done using this rescaling  of $Q_0$  at fixed $x$. 

\section{Comparison with nuclear DIS data, comments and conclusions}

The knowledge of structure functions and parton distribution functions  in nuclei is important in relativistic heavy ion collisions,  since the 
\textquotedblleft hard probes\textquotedblright
 of the quark-gluon plasma require a  control of  cold nuclear effects, i.e.   the   modifications 
depending on the  nuclear dynamics \cite{eskola1,salgado1,yellow}. 

\begin{figure}[h]
\includegraphics[width = 0.35\textwidth]{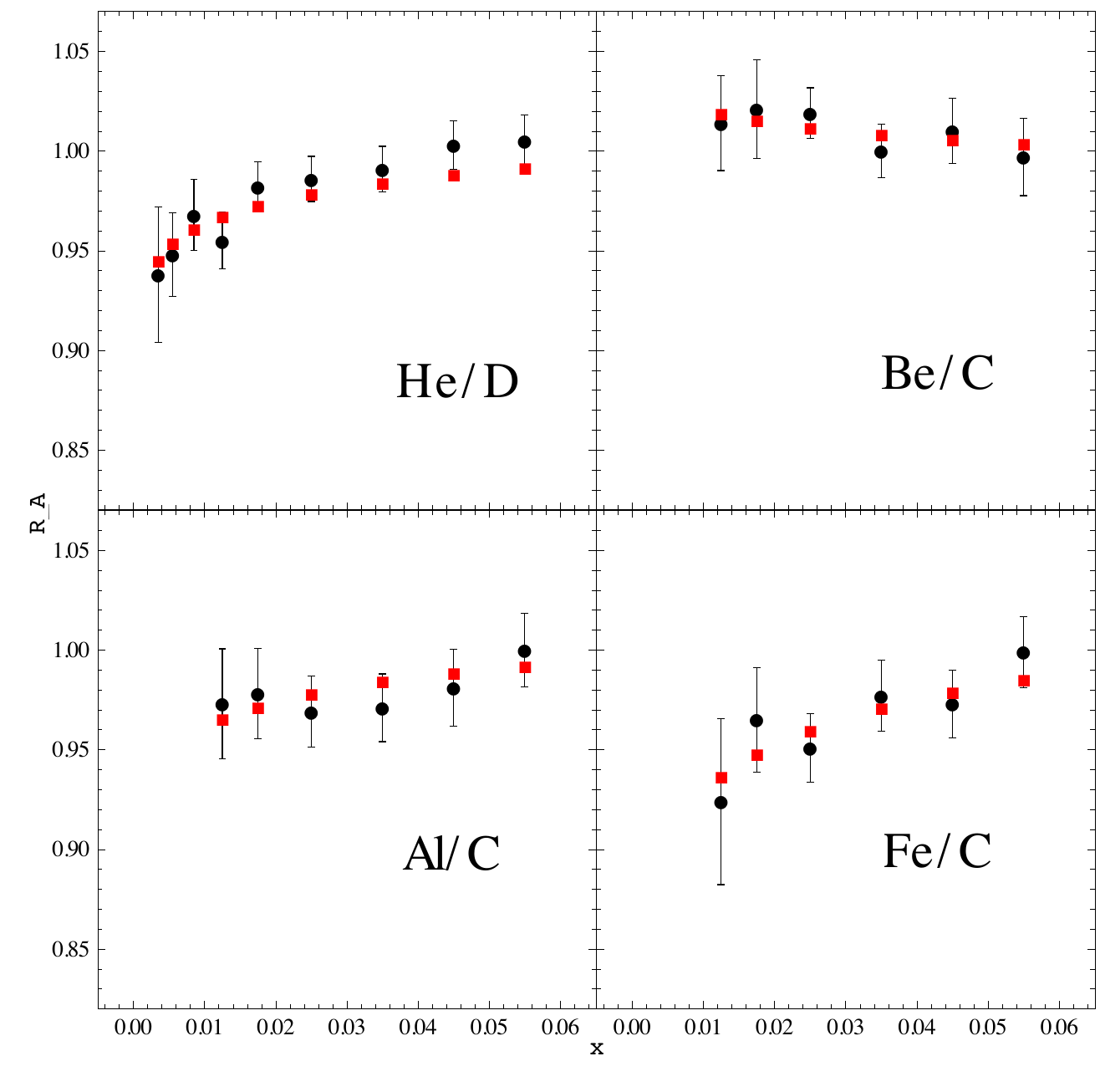}
\caption{Experimental data (black points) \cite{NMC}  and AdS/CFT results based on $Q^2$-rescaling (red squares) for various nuclei.  
 The vaues of the experimental average $Q^2$  (in GeV$^2$), from the first to the last bin in $x$,  vary in the ranges 
$[0.77-6.3]$ (He/D),
$[3.4-9.8]$ (Be/C),
$[3.4-11.6]$ (Al/C),
$[3.4-11.8]$ (Fe/C).
The theoretical values are obtained by Eq.(\ref{F2A})  using the experimental average $Q^2$  for  given $x$ and the 
$\lambda_A$ parameters in the first column of Table \ref{tab:HeD}. 
  The $\chi^2/d.o.f.$ is 1.09 (He/D), 0.21 (Be/C), 0.23 (Al/C) and  0.41 (Fe/C).}\label{fig1}
\end{figure}

\begin{figure}[h]
\includegraphics[width = 0.35\textwidth]{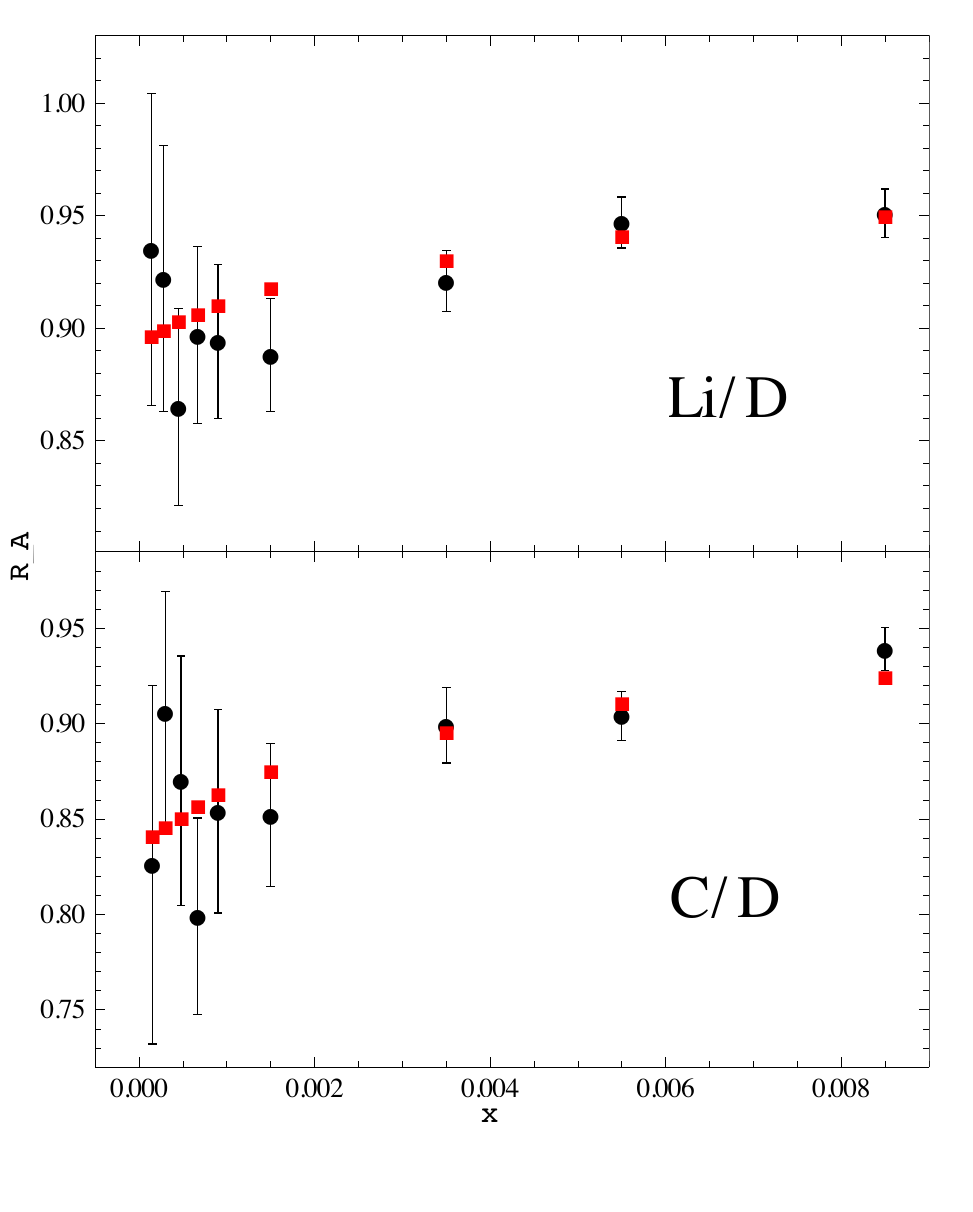}
\caption{ Low-$x$ experimental data for $R_A$ (black points) \cite{NMC}  compared to  holographic results  (red squares) for different nuclei. 
 The experimental average $Q^2$  (in GeV$^2$), from the first to the last bin in $x$, varies in the ranges 
$[0.034-1.4]$ (Li/D),
$[0.035-1.6]$ (C/D).
The theoretical values are obtained  using  Eq.(\ref{F2A}) and the experimental average $Q^2$ for given $x$.
The $\chi^2/d.o.f.$ is 0.93 (Li/D) and  1.61 (C/D).}\label{fig2}
\end{figure}

The results of the holographic  approach  can be compared to experimental data \cite{NMC} by applying the  $Q^2$ rescaling scheme based on Eqs.(12,18): when $Q^\prime_A = \lambda_A  Q^\prime$  one rescales $Q^2 \rightarrow Q^2/\lambda_A^2$ and $Q_0^2 \rightarrow Q_0^2/\lambda_A^2$. The $Q^2$ rescaling is exact not only in the conformal part but also in the confinement contribution. This implies 
\be
F_2^A(x,Q^2) = F^N_{2,cl}\left(x,\frac{Q^2}{\lambda_A^2}\right) + F^N_{2,ct}\left(x,\frac{Q^2}{\lambda_A^2},\frac{Q_0^2}{\lambda_A^2}\right) \,\, , \label{F2A}
\ee
with the conformal term $F^N_{2,cl}$  in Eq.(\ref{F2Nconf1}) and the confinement one $F^N_{2,ct}$  in Eq.(\ref{F2NQ0}). For each nucleus there is only one  parameter $\lambda_A$.

Fig.\ref{fig1} and \ref{fig2} report the comparison of the theoretical results with NMC data \cite{NMC} in the small-$x$ region, $x \leq 0.07$, for different nuclei. The theoretical values are obtained using the average $Q^2$ for a given $x$, the other parameters being fixed as in \cite{Brower:2010wf}. The agreement with data is remarkable, despite the neglect of the  proton and neutron structure function difference. The optimization of the parameters $\lambda_A$ and a complete analysis is presented elsewhere \cite{next};  considering the neutron-proton isospin breaking improves the $\chi^2/ d.o.f.$

Understanding the agreement of the holographic result with nuclear data is easier if we consider 
the origin of  (\ref{Qresc}) and (\ref{Q0resc}).  In the AdS/CFT framework, this  comes from the identification of  the bulk coordinate with the energy scale of  the dual theory: 
considering the form of the $AdS$ metric in Poincar\'e coordinates, a coordinate rescaling $x_\mu \rightarrow \lambda x_\mu$ on the boundary corresponds to $z \rightarrow \lambda z$ in the bulk.
In nuclei, due to the nucleon overlap,  the average distance among quarks and gluons decreases and the color neutralization infrared (confinement) scale increases. 
Such modifications in the boundary correspond in the bulk, respectively,  to $z^\prime \rightarrow z^\prime/\lambda$ and $z_0 \rightarrow \lambda z_0$, i.e.  the prescription employed to describe the nuclear effects by the momenta redefinition. The dynamical generation of the effective IR scale remains to be clarified, with a  possible  analogy with the generation of the saturation scale in free nucleon within this framework \cite{saturation}.

The method inspired by AdS/CFT  provides  a description of the nuclear effects in the DIS structure functions based on $Q^2$-rescaling which corresponds to a geometrical scaling of the confinement size for a bound nucleon.
The same result can be obtained in other, rather different, frameworks. Indeed, at high energy, the structure functions can be
evaluated in the QCD dipole model \cite{dipole1,dipole2} where the virtual photon $\gamma^*$ splits in a quark-antiquark dipole interacting with the target ($T$). 
\begin{table}[b]
	\centering
	\begin{tabular}{l c c c}
	  A & \,\,\,$\lambda_A$(AdS/CFT) \,\,\,& \,\,\,$\lambda_{A,dip}$\,\cite{albacete}\,\,\,&\,\,\, $\lambda_{A,nuc}$\,\cite{close2} \,\,\,\\
	\hline    
Li      &       $1.069$	&   $1.125$ &    	$1.045$\\  
Be	&	$1.073$	&   $1.128$ &    	$1.077$\\
C	&	$1.111$	&   $1.143$ &	        $1.104$\\
Al	&	$1.184$	&   $1.243$ &	        $1.140$\\
Ca	&	$1.219$	&   $1.315$ &	        $1.137$\\
Fe	&	$1.251$	&   $1.387$ &	        $1.154$\\
Pb 	&	$1.327$	&   $1.755$ & 	        $1.188$\\
  \end{tabular}
\caption{$\lambda_A$ obtained  using  $F_2$ in the  holographic approach, compared to the QCD dipole picture 
\cite{albacete} and the change in the nucleus confinement size due to two-nucleon overlap \cite{close2}. }\label{tab:HeD}
\end{table}
Assuming, within this model,  that
the energy and target size dependence of the dipole-target cross section
 $\sigma^{\gamma^* T}$ can be encoded in a saturation scale $Q_{S,T}(x)$ 
\cite{albacete},
that there is no dependence of the dipole
wave function on the quark and antiquarks distribution of the longitudinal
momenta \cite{Schildknecht:2012dc}, and
that there is a minor dependence of the saturation scale on the
specific scattering process,    the dimensionless ratio $\sigma^{\gamma^* T}/\pi R_T^2$
depends only on  $\tau^2_T = Q^2/Q^2_{S,T}(x)$.  The geometric scaling  between the nucleus and the nucleon cross sections  \cite{albacete}
\be
\frac{\sigma^{\gamma^* A}(\tau_A)}{\pi R_A^2}=\frac{\sigma^{\gamma^* N}(\tau_N)}{\pi R_N^2} \,\,\, ,
\ee
with radii $R_{N,A}$ and
\be
\tau^2_A=\tau^2_N\left(\frac{\pi R_A^2}{A \pi R_N^2}\right)^{1/\delta},
\ee
 implies the relation
\be
Q^2_{S,A} = Q^2_{S,N}\left(\frac{A \pi R_N^2}{\pi R_A^2}\right)^{1/\delta}.
\ee
The relation between the structure function per nucleon $F_2^A$ and 
 $\sigma^{\gamma^* A}/\pi R^2_A$ involves the factor $Q^2 \pi R^2_A/A$,  which can be approximated by $Q^2 /Q^2_{S,A}$ if the $x$-dependence of the saturation scale is neglected.
As a result, the dependence of  $F_2^A$  on  $Q^2/Q^2_{S,A}$   
corresponds to rescaling 
$Q^2 \rightarrow Q^2/\lambda^2_{A,dip} ,$ 
with
\be
\lambda_{A,dip} = \left(\frac{A \pi R_N^2}{\pi R_A^2}\right)^{1/2\delta}.
\ee
A good fit of low-$x$ nuclear data in the dipole model is obtained for $R_A = (1.12 A^{1/3}-0.86 A^{-1/3})$ fm, $\pi R^2_N = 1.55$ fm$^2$ and $\delta = 0.79$ \cite{albacete}.

In Table \ref{tab:HeD} we compare the  parameters $\lambda_A$ obtained in  the holographic framework  and in QCD dipole picture. Although the  theoretical approaches are  different, the deviation in this parameter is restrained in the range $5\%-11\%$ going from Li to Fe;  for  Pb  the deviation is about $32\%$.   

The rescaling $Q_0 \to Q_0/\lambda_A$ corresponds to the modification
$z_0^A = \lambda_A z_0$, that is to an increase of the confinement size in nuclei.  
It is interesting to compare the results also with the changes in the confinement size in a nucleus $\lambda_{A,nuc}$ evaluated by the overlap of two nucleons interacting by a Reid soft-core potential \cite{close2}, reported in Table \ref{tab:HeD}. In this case the deviation on $\lambda_A$ is between $4\%$ and  $8\%$ from Li to Fe, and  $12\%$  for  Pb.

 The modification of the confinement sizes is a dynamical property of the
bound nucleons, independent of $x$, which permits the description of
nuclear effects at low-$x$. On the other hand, at larger $x$ one has the
EMC and anti-shadowing regions. A general approach in the whole
kinematical region $0<x<1$ is still lacking, but a $Q^2$ rescaling,
although with different features for large and low-$x$, could be the
unifying dynamical element. We have found that,
starting from  a gauge-gravity duality approach,  a reliable description of nuclear shadowing can be obtained rescaling the  virtual photon scale $Q^2$ \textquotedblleft seen\textquotedblright   by a parton in a bound nucleon.

\end{document}